\pdfoutput=1
\documentclass[a4paper,10pt,amsmath,amssymb,nofootinbib,onecolumn,
floats,longbibliography,aps,prd]{revtex4-2}
\usepackage[british]{babel}
\usepackage{mathrsfs,bm}
\usepackage{natbib,textcase}
\usepackage{graphicx,color,soul,comment}
\usepackage[pdftex,pdfusetitle]{hyperref}
\hypersetup{colorlinks=true,linkcolor=red,citecolor=blue,filecolor=black,urlcolor=black,
pdfauthor={Vania Vellucci, Edgardo Franzin and Stefano Liberati}}
\usepackage[capitalize]{cleveref}
\crefname{subequation}{Eqs.}{Eqs.}
\labelcrefformat{subequation}{#2(#1)#3}
\usepackage[caption=false]{subfig}
\usepackage{physics}
\usepackage{array,booktabs} 
\usepackage{multirow} 
\newcolumntype{L}{>{$}p{20mm}<{$}} 
\makeatletter\g@addto@macro\bfseries{\boldmath}\makeatother%

\allowdisplaybreaks

\def\be#1\ee{\begin{align}#1\end{align}}

\newcommand{\ie}{i.e.}
\newcommand{\eg}{e.g.}
\newcommand{\cmp}{\sigma}
\renewcommand{\dd}{\text{d}}

\newcommand{\kabs}{\kappa}
\newcommand{\p}{\partial}
\newcommand{\0}{\nonumber}

\renewcommand{\geq}{\geqslant}

\DeclareMathOperator\const{const}

\begin{document}

\title{Echoes from backreacting exotic compact objects}
\date{\today}

\newcommand{\SISSA}{\affiliation{SISSA, International School for Advanced Studies, via Bonomea 265, 34136 Trieste, Italy}}
\newcommand{\InfnTS}{\affiliation{INFN, Sezione di Trieste, via Valerio 2, 34127 Trieste, Italy}}
\newcommand{\IFPU}{\affiliation{IFPU, Institute for Fundamental Physics of the Universe, via Beirut 2, 34014 Trieste, Italy}}

\author{Vania Vellucci}
\author{Edgardo Franzin}
\author{Stefano Liberati}
\SISSA\IFPU\InfnTS

\begin{abstract}
The possible detection of echoes in late gravitational-wave signals is the most promising way to test horizonless alternatives to general relativistic black holes, and probe the physics of these hypothetical ultra-compact objects.
While there is currently no evidence for the presence of such signatures, better accuracy is expected with the growing wealth of data from gravitational waves observatories. 
So far, several searches for these specific signals have been performed considering equidistant intervals between consecutive echoes, \ie\ quasi-periodic wave-forms, and ignoring possible backreaction effects of the incoming waves.
Here we study scalar perturbations in exotic compact object scenarios that account for possible backreaction phenomena.
In particular, we find that if one considers the increase of the central object mass due to  the partial absorption of the energy carried by the perturbation, the echo signal can be quite different and non-periodic.
Apart from this simple scenario, we also consider the case in which, in order to preserve its compactness above the black hole limit, the compact object absorption shuts down in a finite amount of time or leads to an expansion.
In both these cases we find interesting new features that should be taken into account in future searches.
\end{abstract}

\maketitle%

\section{Introduction}

Horizonless black-hole mimickers are foreseen from quantum gravity theories (\eg\ fuzzballs~\cite{Mathur:2005zp}, gravastars~\cite{Mazur:2004fk, Uchikata:2015yma}) and beyond standard model physics (\eg\ boson stars~\cite{Liebling:2012fv}).
Being able to distinguish such exotic compact objects (ECOs) from classical black holes (BHs) and to obtain information about their structure would hence allow us to probe new physics and possibly give us hints about the quantum gravitational effects behind the regularization of general relativistic singularities.

Presently, the most powerful way to do so, is through gravitational wave (GW) signals from the coalescence of BH candidates.
In particular, a proof of the presence/absence of an event horizon could come from the ringdown phase of these events.
This phase is caused by the oscillations of the final perturbed object formed in the coalescence and it is governed by a series of damped oscillatory modes.
In the high-compactness limit, the ringdown signal of an ECO is initially almost identical to that of a BH with the same mass, but then it is followed by a series of secondary pulses~\cite{Cardoso:2016rao,Cardoso:2016oxy}.

To understand the origin of these pulses it is sufficient to analyze the field equation of a  perturbation in the BH and ECO spacetimes, characterized by the same Arnowitt--Deser--Misner (ADM) mass $M$.
Consider for definiteness spherical symmetry. In both cases, the wave encounters a potential barrier peaked at the photon sphere $r\approx 3M$.
For a BH spacetime, the event horizon behaves as a perfectly absorbing surface, while for an ECO spacetime, the key difference is the presence of another barrier at the location of its surface $r_0$, which can in principle reflect incoming radiation. 
The presence of this cavity between the ECO surface and the photon sphere can create quasi-trapped modes that can tunnel through the photon sphere potential and travel towards infinity in the form of secondary pulses with smaller and smaller amplitude and frequencies. For these reasons, these subsequent and similar signals are called \emph{echoes}~\cite{Cardoso:2017cqb,Cardoso:2019rvt,Annulli:2021ccn}.

Usually echoes are studied in linear perturbation theory, neglecting the possible backreaction of the ECO.
This is justified because the energy in the ringdown is small (from two to three orders of magnitude with respect to the mass of the object~\cite{Baibhav:2017jhs,Barausse:2012qz}) and diffused on a wavelength of the order or larger than the ECO radius.
However, it can be argued that this linear approximation is actually dangerous since for good BH mimickers the surface of the ECO is very close to the would-be horizon~\cite{Carballo-Rubio:2018jzw}. Indeed, the peeling of geodesics will cause an accumulation of light rays and so a large increase of the perturbation energy density near the surface. Actually, it can be shown that GW fluxes can even lead to the violation of the hoop conjecture~\cite{Thorne:1972} and the collapse of the ECO into a BH~\cite{Carballo-Rubio:2018vin,Addazi:2019bjz,Chen:2019hfg}.
Thus it seems clear that non-linear interactions between the ECO and the GW flux must be considered.

The scope of this work is to study possible effects on the echoes waveform due to backreaction and in particular we shall focus on the effect of the echo absorption on behalf of the ECO.
A perfectly reflecting surface is in fact just an idealization, we expect instead that part of the radiation can be absorbed by the object through effects of dissipation or viscosity~\cite{Esposito:1971}.
The introduction of an absorption coefficient for the ECO surface is not only physically reasonable, but it also allows to circumvent some instabilities of ECOs connected to the presence of a stable light ring~\cite{Cunha:2017qtt,Ghosh:2021txu} and of an ergoregion~\cite{Cardoso:2007az,Pani:2010jz,Barausse:2018vdb}.
Indeed, it makes the timescale for the onset of these instabilities very long and it can even turn it off~\cite{Maggio:2017ivp,Maggio:2018ivz}. 

However, until now, the only considered effect of absorption on the echoes waveform is the decreasing of echoes amplitude and energy~\cite{Maggio:2018ivz,Maggio:2019zyv} or the changing of their frequency content in the case of frequency-dependent absorption coefficient~\cite{Oshita:2019sat,Chakraborty:2022zlq}.
Yet, it should also imply an increase of the central object ADM mass and consequently a change of the spacetime in which echoes are propagating. 
Similar considerations were done in Ref.~\cite{Sberna:2021eui} in the context of BH ringdown, showing how the change in the BH mass, due to the absorption of a mode excited at early times, causes a shift in the mode spectrum and thus the excitation of additional modes.

In the first part of this work, we consider the simple instructive case in which we have no other backreaction effects apart from absorption.
We show that this leads to a non-constant time delay between echoes and thus the loss of the typical quasi-periodicity predicted for these signals. 
This is particularly interesting if we think that the strategies used in the searches for echoes in the GW ringdown~\cite{Abedi:2016hgu,Ren:2021xbe,Abedi:2020sgg} are usually based on the aforementioned quasi-periodicity. 
Indeed, in Ref.~\cite{Wang:2019szm} it was shown that applying a template with constant time interval between echoes may significantly misinterpret the signals if the variation of this interval is greater than the statistical errors.

In the second part of this investigation, we take into account that, for sufficiently compact central objects, the absorption of part of the GW flux can increase the mass of the object over the hoop limit $2M \geq r_0$. Thus, assuming the stability of these ECOs, some backreaction mechanism must be present so to prevent the formation of a horizon: We consider a scenario in which the absorption coefficient depends on the compactness, and a scenario in which the ECO expands.
In both cases we show the effects on the echoes waveform and time delay.

\section{Set-up}

As a proxy to the more general case of gravitational perturbations, here we study the evolution of a minimally coupled massless scalar field $\Phi$ propagating in a spherically symmetric ECO spacetime.
As commonly assumed, the scalar field does not couple directly to any form of matter that might be present within or outside the ECO.
Its action reads
\be\label{scalaraction}
S=-\frac{1}{2}\int \dd^4 x \sqrt{-g}\,g^{\mu\nu} \nabla_\mu\Phi \nabla_\nu\Phi\,.
\ee

We do not assume any specific form for the gravitational action and the ECO spacetime is characterized by its initial ADM mass $M_0$ and radius $r_0$ greater than its Schwarzschild radius.
We define the compactness parameter of the object as
\be\label{defcompactness}
\cmp\equiv\frac{r_0}{2M_0}-1\,,
\ee
which is always positive and goes to zero in the BH limit, \ie\ $\cmp\to0$ as $r_0\to2M_0$.
Independently of the specific ECO model, the spacetime outside its surface is Schwarzschild. For $r>r_0$ we have:
\be
\dd s^2=-f(r)\,\dd t^2+\frac{\dd r^2}{f(r)} + r^2\left(\dd\theta^2 + \sin^2\theta\,\dd\varphi^2\right),
\quad f(r) = 1 - \frac{2M_0}{r}\,.
\ee

Because of spherical symmetry, the Klein--Gordon equation in this spacetime is separable thus we can decompose the field as $\Phi(t,r,\theta,\varphi)=\sum_{lm}\frac{\Psi_{lm}(t,r) }{r} Y_{lm}(\theta,\varphi)$, where $Y_{lm}$ are the scalar spherical harmonics.
Then the field equation for each mode $\Psi_{lm}(t,r)$ is (to avoid cluttering, in what follows we drop the $lm$ indexes):
\be
\frac{\p^2 \Psi}{\p t^2} - f^2 \frac{\p^2 \Psi}{\p r^2} - f f' \frac{\p \Psi}{\p r}+V(r) \Psi=0\,,\label{radialKG}
\ee
where a prime represents derivative with respect to the radial coordinate $r$ and 
\be
V(r)=\left(1-\frac{2M_0}{r}\right)\left(\frac{2M_0}{r^3}+\frac{l(l+1)}{r^2}\right).
\ee

\subsection{Energy of the perturbation}

To compute the energy of the scalar perturbation, we start with the stress-energy tensor stemming from the action in \cref{scalaraction}
\be
T^{\mu\nu} = \nabla^\mu\Phi \nabla^\nu\Phi - \frac{1}{2} g^{\mu\nu}\nabla_\alpha\Phi \nabla^\alpha\Phi\,.
\ee
We consider the conserved current $J^{\mu}=k_{\nu} T^{\mu \nu}$, where $k_\nu$ is the timelike Killing field of the Schwarzschild spacetime.
The conserved energy in a three-dimensional hypersurface $\Sigma$ is then
\be
E = \int_\Sigma \dd^3x \sqrt{\gamma}\, J^{\mu}n_{\mu}\,,
\label{E}
\ee
where $\gamma$ is the induced three-dimensional metric on the hypersurface and $n^\mu$ is the normalized vector field orthogonal to $\Sigma$. In our case $\Sigma$ is the hypersurface at $t=\const$ and thus $n^{\mu}=\nabla ^{\mu}t/|\nabla^{\mu}t| = k^{\mu}/\sqrt{f(r)}$.
From this, integrating in the angular part, we obtain, \eg\ in Schwarzschild coordinates
\be
E=\frac{1}{2} \int \frac{\dd r}{f(r)}\left[ \left(\frac{\p\Psi}{\p t}\right)^2 + f(r)^2 \left(\frac{\p\Psi}{\p r}\right)^2 +  V(r)\,\Psi^2 - f(r) \frac{\p}{\p r} \left(\frac{f(r)}{r}\Psi^2 \right) \right].
\label{Es}
\ee
These coordinates might look not suitable, as the factor $1/f(r)$ gets divergent as we approach the event horizon.
However, the field $\Psi$ --- seen by a static observer at infinity --- moves slower and contracts as it approaches the horizon, reaching it in an infinite amount of time, while the region $\Delta r$ in which the field is diffused shrinks.
The two effects compensate each other, and the energy remains constant.

\subsection{Varying mass and moving surface}

If the object absorbs energy from the scalar field, we have to take into account that its mass can change in time.
We assume that at each instant the spacetime can be described by the Schwarzschild metric with a different mass $M_0\to M(t)$ and $f(r)\to F(t,r)=1-2M(t)/r$.
Then the Klein--Gordon equation, written in terms of the ``initial'' tortoise coordinate $r_*=r+2 M_0 \ln \left(r/2 M_0 -1\right)$, becomes:
\be\label{radialKG_varyingM}
\frac{\p^2 \Psi}{\p t^2} - \frac{F^2}{f^2} \frac{\p^2 \Psi}{\p r_*^2} - \frac{F}{f}\left(\frac{\p F}{\p r}-\frac{F f'}{f} \right) \frac{\p \Psi}{\p r_*} + V(r,t) \Psi + \frac{1}{F}\frac{\p F}{\p t}\frac{\p \Psi}{\p t}=0\,.
\ee
where the potential $V$ has been promoted as a function of $t$ and $r$, and $r$ itself is interpreted as a function of $r_*$.

In one of the models that we consider the surface of the object moves, and thus the point at which we impose our boundary conditions, \ie\ $r_0 \to r_0(t)$.
Thus to solve the scalar field equation with a time-independent boundary condition we need to choose a coordinate in which the surface of the object is fixed in time.
For example, if the object expand in order to stay at constant compactness then $x=M_0 \ln \left(r/2M(t) -1\right)$ is a good choice.
In these coordinates the Klein--Gordon equation reads
\be\label{KG_movingsurface}
&\frac{\p^2 \Psi}{\p t^2}-M_0^2\left( \frac{1}{r^2}-\frac{r^2 \Dot{M}^2}{M^2(r-2M)^2} \right)\frac{\p^2 \Psi}{\p x^2} - \frac{2\Dot{M} M_0 r}{M (r- 2M)} \frac{\p^2 \Psi}{\p t \p x}\0\\
&+ M_0 \left( \frac{1}{r^2}-\frac{2 M}{r^3}-\frac{r \Ddot{M}}{M (r- 2 M)} +\frac{r (r - 4 M) \Dot{M}^2}{M^2 (r-2M)^2} \right) \frac{\p \Psi}{\p x} + V(r,t) \Psi=0\,,
\ee
where the time dependence of $M$ is implicit, and $r$ is interpreted as a function of $x$.

In these coordinates the surface, at which we impose the boundary condition, is always at $x = x_0 =M_0\ln\left(r_0/2 M_0 - 1\right)$.
Another way to simulate a  moving surface is to simply change, at each time step, the point of the numerical grid at which we impose the boundary condition, the two methods bring to the same results.

\subsection{Numerical set-up}

In the numerical simulation reported in the next sections we always consider an $l=2$ quadrupolar mode and we use as initial condition for $\Psi$ a Gaussian pulse:
\be
\frac{\p \Psi(r,0)}{\p t}= \Psi_0 \exp\left(-\frac{(r_*-r_c)^2}{2s^2}\right), \quad \Psi(r,0)=0\,,
\ee
with central value $r_c=11 M_0$ and width $s=2 M_0$; different initial values lead to similar results.
The pulse is initially centered outside the potential barrier $V(r)$, whose peak is at approximately $3M_0$, and it is moving inward.
The amplitude $\Psi_0$ is chosen in order to obtain an impulse with energy of order of the one we expect to be contained in the echoes signals, roughly from two to four orders of magnitude smaller than the mass of the central object~\cite{Abedi:2016hgu}.

We evolve $\Psi$ in the time domain using a fourth-order Runge--Kutta integrator and computing spatial derivatives with finite-difference approximation of second-order accuracy~\cite{Dolan:2011dx}. 
A convergence test of the code is shown in \cref{fig:conv} and discussed in \cref{app:numerical}.
The non-trivial boundary conditions that we have imposed for our numerical simulations are described in \cref{app:boundary}.

During the simulations the mass of the central object increases because of the energy absorbed from the field, at a given time step, by an amount
\be
\kabs\,\Delta E(t)\,,
\ee
where $\kabs$ is the absorption coefficient (see definition below) and $\Delta E(t)$ is the field energy present, at time $t$, in the last spatial bin of our computational domain corresponding to the location of the surface of the central object.
The method with which we estimate $\Delta E(t)$ at each instant is explained in \cref{app:absorbed}.

\section{Echoes: absorption beyond the test field limit\label{s:absorption}}

In our first scenario, we evolve $\Psi$ according to \cref{radialKG_varyingM}, taking into account that the mass of the compact object can increase during the evolution as a consequence of the energy absorbed from the scalar field.
In fact, whatever is the mechanism responsible for absorption, the energy of the field is converted in some other kind of energy inside the object, \eg\ thermal energy in the case of dissipative/viscous effects, and since all energy  ``gravitates'', this absorption will increase the ADM mass of the object.

We define the absorption coefficient as the fraction of the incoming energy $E_\text{in}$ that is lost inside the object:
\be\label{kabsorption}
\kabs = 1 - \frac{E_\text{out}}{E_\text{in}}\,.
\ee

Given a certain compactness and absorption coefficient, there exists a maximum flux of energy that the object can sustain without overcoming the hoop limit and collapse into a BH.
It follows that if the energy in the GW flux is bigger than this maximum, the collapse will delay part of the echoes signal.
In the examples shown here, we have chosen compactness and absorption coefficients in order to always remain below this limit.

In linear approximation, the time delay between echoes is the time that light takes to travel from the potential barrier centered around $r_\text{peak}\approx 3M_0$ to the ECO surface $r_0=2M_0 (1+\cmp)$ and back,\footnote{In reality, it should be taken into account the interaction time $\Delta t_\text{int}$ during which the field travels inside the interior of the object and it is partially absorbed.
Here, we assume it to be of the order of the radius of the object and thus negligible with respect to $\Delta t_\text{echo}$.
However, there exist models in which the interior spacetime is such that $\Delta t_\text{int}$ becomes dominant~\cite{Pani:2018flj,Raposo:2018rjn}, but it also depends on the compactness parameter. The precise dependence on $\cmp$ is however model-dependent and generically not logarithmic.} as follows
\be
\Delta t_\text{echo} = 2 \int_{r_0}^{r_\text{peak}} \frac{\dd r}{f(r)} \simeq 2M_0 \left[1-2\cmp-2 \ln (2\cmp) \right],
\label{dt}
\ee
where $\cmp$ is the compactness parameter defined in \cref{defcompactness}.

Nonetheless, if the ECO absorbs a small quantity of energy $\Delta E$ from the first echo, increasing its mass as $M_0\to M = M_0 + \Delta E$ but remaining with the same radius $r_0$, the ECO compactness parameter for the first and second echoes is different %
\be\label{DeltaEchoes}
\cmp_\text{1st echo} = \frac{r_0}{2M_0}-1\,,\quad
\cmp_\text{2nd echo} = \frac{r_0}{2 M}-1=\frac{r_0}{2(M_0+\Delta E)}-1 < \cmp_\text{1st echo}\,.
\ee

\begin{figure}[ht]
\centering
\includegraphics[width=0.5\textwidth]{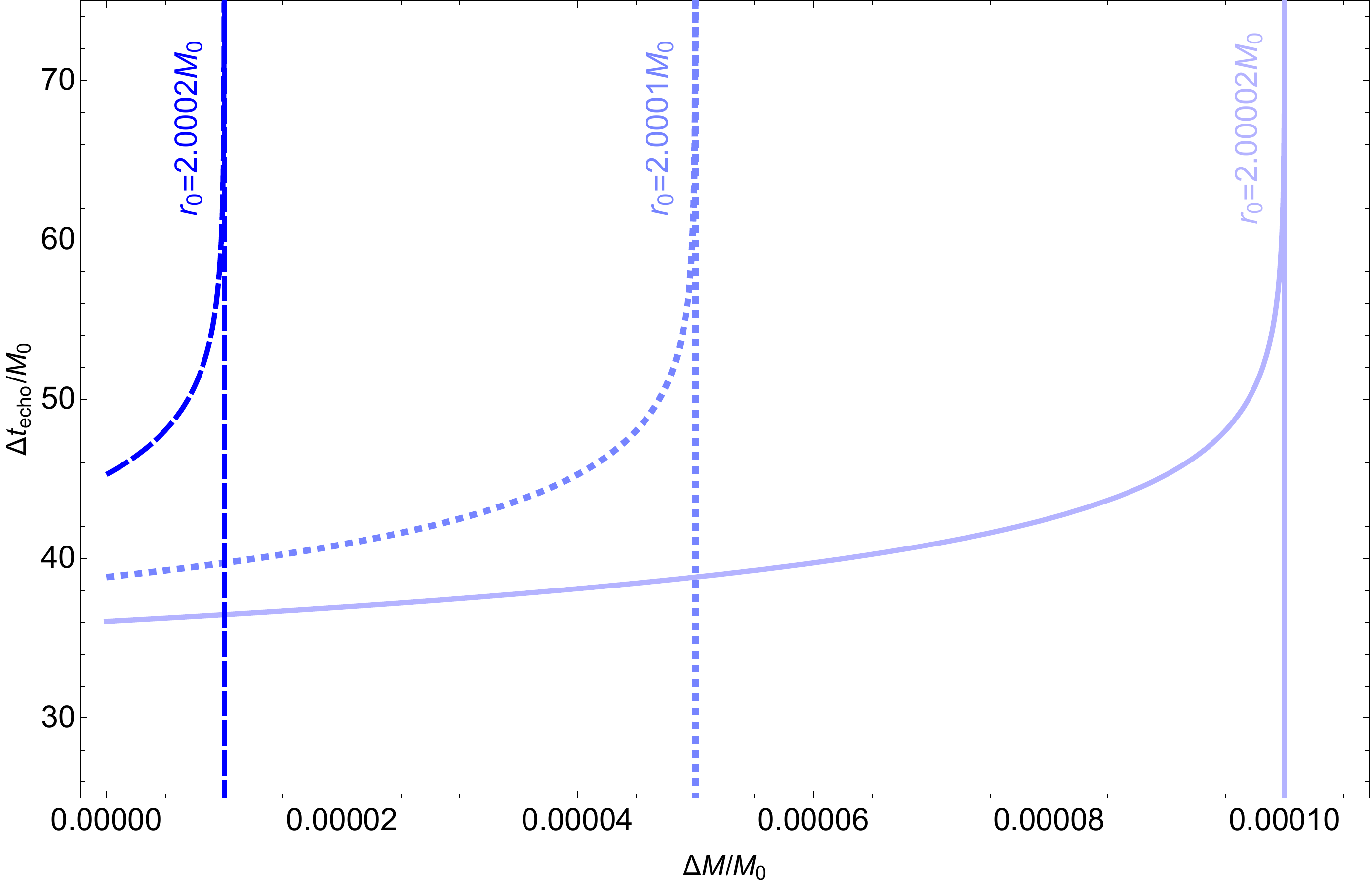}
\caption{Time delay between echoes for fixed position of the surface $r_0$ and variable mass. The vertical lines represent the asymptotic value of $\Delta M/M_0$ for which the BH limit $r_0=2M_0$ is exceeded. All values are reported in units of the initial mass of the object $M_0$.}
\label{fig:dt}
\end{figure}

As a consequence, also the time delay of the second echo will be different.
We give some numerical examples in \cref{fig:dt} and \cref{tab1} for selected values of the compactness and absorption parameters.

In these examples and in the plots that we show across the article, it is clear that even very small absorption coefficients (of order 0.1\%--0.01\%) lead to significant changes of the signal.
This can seem counterintuitive, given the small amount of energy present in the echoes. The point is that, although the change in the mass is actually small, for very compact objects it is sufficient to cause a big change in their compactness if the radius remains fixed.
Consider for definiteness an ECO with initial compactness parameter $\cmp_0 = 10^{-7}$: the absorption of the amount of energy $5\cdot 10^{-8}\,M_0$ is sufficient to halve its compactness parameter and, as a result, to significantly change the spacetime in which the field is propagating.
In this example, the time delay between echoes that depends logarithmic in the compactness parameter, roughly changes from $\Delta t_\text{echo}/M_0 \approx -4 \ln(2\sigma_0) \approx 61.7$ to $\Delta t_\text{echo}/M_0 \approx 64.5$.

\begin{table}[ht]
\begin{tabular}{LLLLLLL}%
\toprule
\cmp_0 &  \Delta t_{\text{echo}}/M_0 &\kabs & E_\text{1st echo}/M_0 & \Delta t_1/M_0 & E_\text{2nd echo}/M_0 &  \Delta t_2/M_0 \\ \midrule
 10^{-3} &     26.85   & 6\% & 10^{-2}  &   30.54   & 0.25 \cdot 10^{-2}  & 32.43\\ \hline
 
    10^{-4} &    36.07    &  6\% & 10^{-3} &   39.74  &  0.25 \cdot 10^{-3}    &  41.62   \\ \hline
10^{-5}      &  45.28   &  6\%  &  10^{-4}    &    48.94  &   0.25 \cdot 10^{-4} & 50.82 \\ \hline
   10^{-5}  &  45.28    & 0.06\%     &   10^{-2}    &   48.94    &   0.25 \cdot 10^{-2} & 50.82 \\ \hline
   10^{-6}&      54.49     &  0.06\%  &    10^{-3}  &  58.15   &   0.25 \cdot 10^{-3}       &   60.03              \\ \hline
      10^{-7}&     63.70    & 0.06\%  &    10^{-4}  & 67.36  &   0.25 \cdot 10^{-4}       &          69.24      \\
\bottomrule
\end{tabular}

\caption{Given an initial compactness parameter $\cmp_0$  we report the expected time delays between echoes $\Delta t_\text{echo}$  in the case of a perfect reflecting ECO and the true time delay between the first and the second echo $\Delta t_1$ and between the second and third echo $\Delta t_2$ in the case of an ECO with absorption coefficient $\kabs$. These last two time delays are different to respect to $\Delta t_\text{echo}$ because the partial absorption of the energy contained in the first and second echo changes the compactness of the central object.  We assume that the energy in the second echo is approximately a quarter of that of the first one. All values are reported in units of the initial mass of the object $M_0$. }\label{tab1}
\end{table}

\begin{figure*}[ht]
\centering
\includegraphics[width=0.7\textwidth]{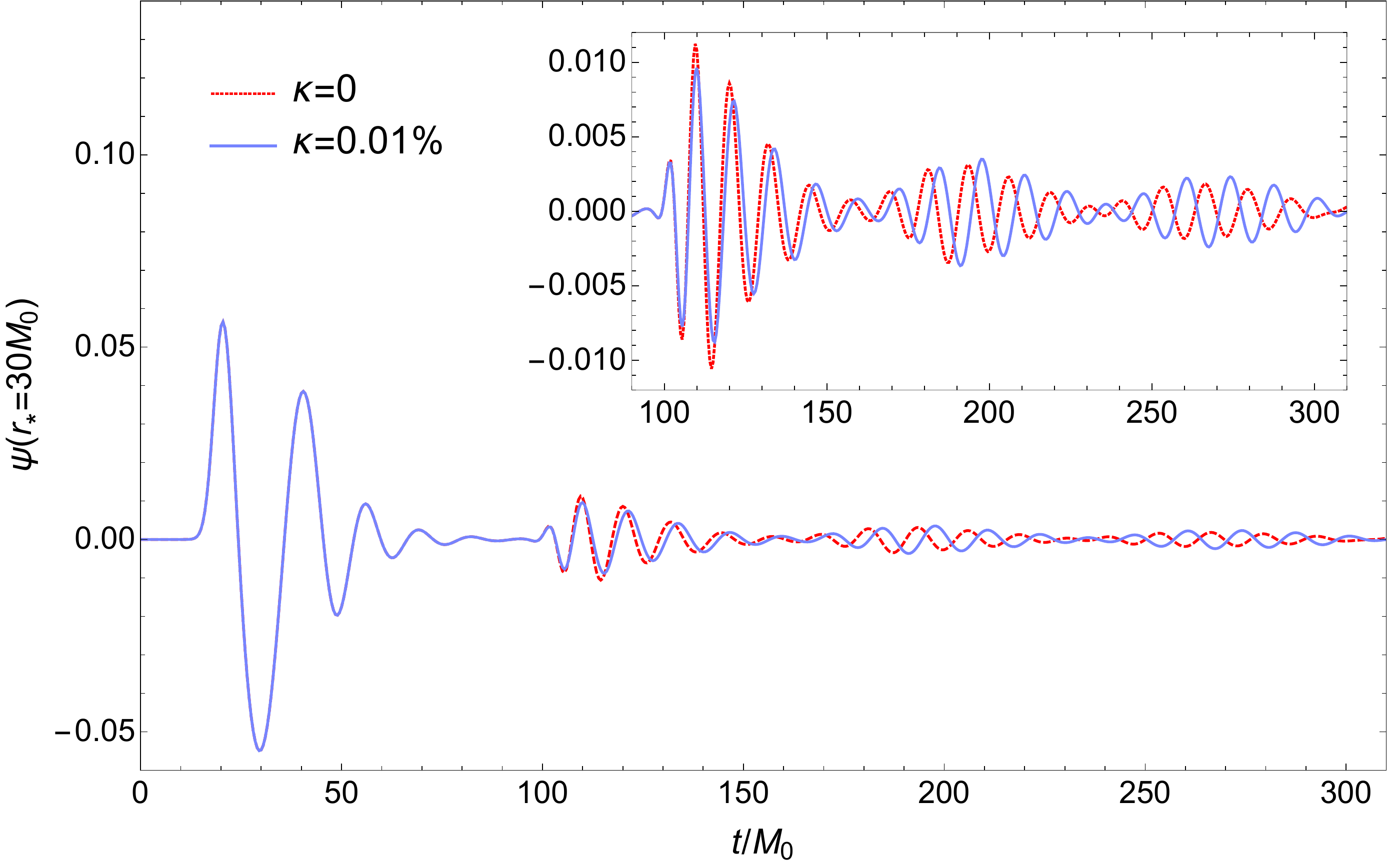}
\caption{Quadrupolar waveform extracted at $r_*=30 M_0$. The initial Gaussian pulse has energy $E=4.4 \cdot 10^{-3} M_0$ and it is scattered off an ECO with initial compactness parameter $\cmp=10^{-7}$, \ie\ $r_0=2.0000002 M_0$.
The red dashed line shows the case of a perfectly reflecting object, while the purple solid line shows the case in which the surface of the object absorbs incoming radiation with an absorption efficiency of $0.01\%$.}
\label{dM0000000055}
\end{figure*}

In \cref{dM0000000055} we plot the typical waveform for a Gaussian pulse scattered off an ECO with a small absorption parameter, compared with a perfectly reflecting ECO.

We observe a considerable difference in phase between the two cases due to the aforementioned non-negligible change in the time delay among echoes, in agreement with the analytical estimate of \cref{dt}.

We also note a difference in the amplitude of the signal. The absorption coefficient is too small to produce a visible decreasing of the echoes amplitude, yet the first echo has a smaller amplitude with respect to the perfectly reflecting case, while the subsequent echoes have even bigger amplitudes. 
To understand this redistribution of energy is convenient to look at the field equation in $(t,r_t)$ coordinates, where $r_t$ is the ``time-dependent'' tortoise coordinate $r_t = r +2 M(t) \ln\left(r/2 M(t) - 1 \right)$, in which the Klein--Gordon equation remains a simple wave equation apart from some negligible terms proportional to $\dot{M}$ and $\ddot{M}$.
Nevertheless, the position of the surface in the $r_t$ coordinate changes with time and gets more negative as the mass increases, since the object is becoming more compact.
This means that while the first echo is reflecting on the central object surface, this surface is moving away from it. Since we are in a reference frame in which the field equation is a simple wave equation, a movement of the surface causes a Doppler effect, \ie\ a decrease of the frequency content of the field.
For this reason, a smaller fraction of the first echo will pass through the high-pass filter potential centered at $r_\text{peak}$, while a greater fraction of it will be bounced back forming the subsequent echoes that therefore will have a greater amplitude with respect to the perfectly reflecting case.
Also the absorption of the second and third echoes causes an increase of the mass and thus a movement of the surface during the reflection, but the effect is quite negligible with respect to the previous increase of amplitude.

\section{Echoes: absorption and backreaction scenarios}

In realistic situations, the energy absorbed from the ringdown signal might be enough to cause the collapse of the ECO into a BH.
Thus, if we want the object to be stable, we have to take into account some mechanisms that prevent the collapse, whose details depend obviously on the physics of the object and the specific gravitational field equations.
To have a glimpse on the possible effects on the signal, while remaining agnostic about the novel physics supporting the ECO, we consider here some simple model-independent scenarios.

\subsection{Asymptotic compactness \label{s:limitcomp}}

As a first scenario, the object is allowed to absorb energy varying its compactness up to a certain limit, say until its surface is at a Planck length from the would-be horizon.
This corresponds to a compactness parameter of order of $\cmp_\text{Planck}\approx 10^{-40}$ for stellar mass objects ($M\approx 10^2 M_{\odot}$).
This situation is particularly reasonable because the ringdown signal comes from objects that have just been formed in a binary coalescence, thus it is not obvious that they are already in their definitive stable configuration.

Technically, this situation can occur when the ECO absorption coefficient depends on its compactness and goes to zero as $\cmp \to \cmp_\text{Planck}$. 
The effects on the echoes waveform will depend on the initial value of the absorption coefficient, on the velocity with which it goes to zero and on the value of the asymptotic compactness.
If the absorption coefficient varies very slowly, we obtain the same results of constant absorption, analyzed in the previous section, and no other backreaction.
When instead the absorption coefficient goes rapidly to zero, only the first echo will be partially absorbed and the time delay will soon stabilize to a constant value. 
Among several possible choices, in our computations we have considered this functional form for the absorption coefficient
\be\label{k2}
\kabs(\cmp) = \alpha \left(1-\tanh{\frac{\beta}{\cmp-\cmp_\text{Planck}}}\right),
\ee
which varies slowly when the compactness is far from the Planck value, and goes smoothly to zero for $\cmp\to\cmp_\text{Planck}$. The parameters $\alpha$ and $\beta$ represent, respectively, the initial value and the velocity with which $\kabs$ goes to zero.
At each step of the simulation
\be
\cmp(t+\Delta t)=\frac{r_0}{2 M(t)+\kabs(\cmp(t)) \Delta E(t)}-1,
\ee
such that as $\cmp(t)\to\cmp_\text{Planck}$, we have $\kabs(\cmp) \to 0$ and thus $\cmp(t+\Delta t) =\cmp(t)$.

\begin{figure*}[ht]
\centering
\includegraphics[width=0.8\textwidth]{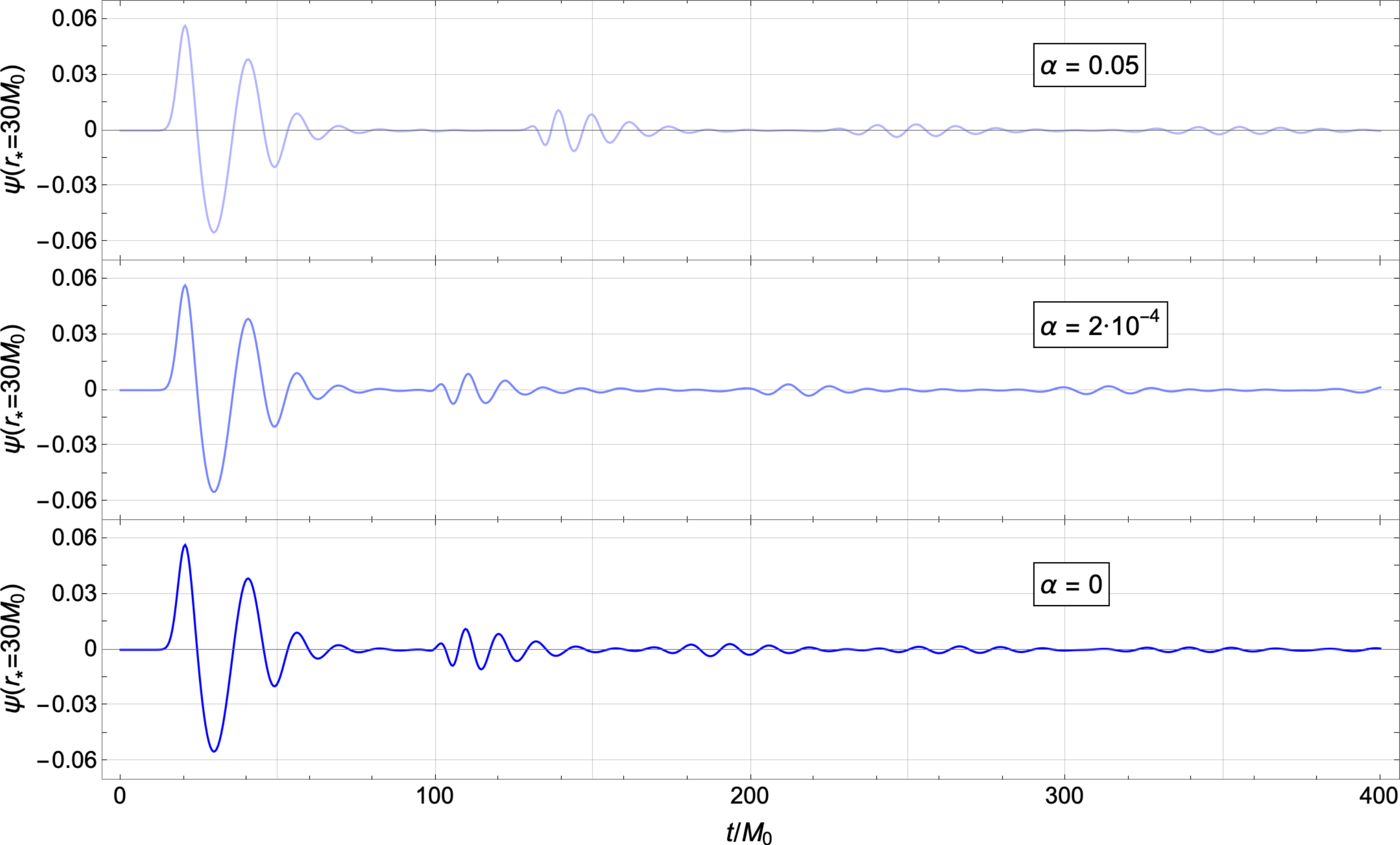}
\caption{Quadrupolar waveform extracted at $r_*=30 M_0$. The initial Gaussian pulse has energy $E=4.4 \cdot 10^{-3} M_0$ and it is scattered off an ECO with initial compactness parameter $\cmp=10^{-7}$, \ie\ $r_0=2.0000002 M_0$. We show results for a compactness-dependent absorption parameter (top and middle panels, for selected values of the parameter $\alpha$ introduced in \cref{k2} and $\beta=\cmp\cdot 10^{-3}$) and for a perfectly reflecting ECO (bottom panel).}
\label{depend}
\end{figure*}

In \cref{depend} we plot the typical waveform for a Gaussian pulse scattered off an ECO with a compactness-depending absorption parameter, compared with a perfectly reflecting ECO.

The top and middle panels show our results for, respectively, $\alpha = \{0.05, 2\cdot 10^{-4}\}$ and $\beta=\cmp_0 \cdot 10^{-3}$, with $\cmp_0=10^{-7}$.
In the first case the initial absorption coefficient is $\kabs_0 \approx 5 \%$. 
With an incoming Gaussian pulse of energy $E\approx 10^{-3}M_0$, this $\kabs_0$ is sufficient to reach a very high compactness, and thus a negligible absorption coefficient, already during the absorption of the very first part of the first echo.
For this reason, we observe that the time delay among the echoes is approximately constant even if it is larger than in the perfectly reflecting case shown in the bottom panel.
In the second case, the absorption coefficient is approximately constant and $\kabs_0\approx 0.02\%$ during the absorption of the first echo, while it becomes negligible only during the absorption of the third echo. This leads to non-constant time delays among the first echoes. 

We want to mention another possible scenario in which the change in the compactness stops when the object reaches a limiting value: the absorption coefficient is constant but, once the asymptotic compactness is reached, the ECO starts expanding to remain in the same equilibrium configuration.
In this case, the first part of the signal will have larger time delays but this increment of $\Delta t_\text{echo}$ will stop once the expansion starts. The reason for this will be clear in the next section.

\subsection{Expansion}

In another backreaction scenario, to remain stable, the ECO compensates the absorption by expanding.
A reasonable way to model the expansion is to assume that its radius moves in order to maintain the same initial compactness, according to the following prescription (in Schwarzschild coordinates)
\be
\frac{r_0(t+\Delta t)}{2\left(M(t)+\Delta E\right)} \overset{!}{=} \frac{r_0(t)}{2M(t)}=\cmp_0+1\,,
\ee
where $\Delta E$ is the amount of energy absorbed from the scalar field in the time interval $\Delta t$. 
  
In this case, the sole dependence on time in \cref{dt} is a linear dependence on $M(t)$ and thus the time delays among echoes will be approximately always the same, as in the perfectly reflecting case.
Obviously, this is true only if the expansion happens fast enough for the object to be approximately at the same constant compactness at each instant --- note that this could even require superluminal expansion~\cite{Carballo-Rubio:2018vin}.

\begin{figure}[ht]
\centering
\includegraphics[width=0.5\textwidth]{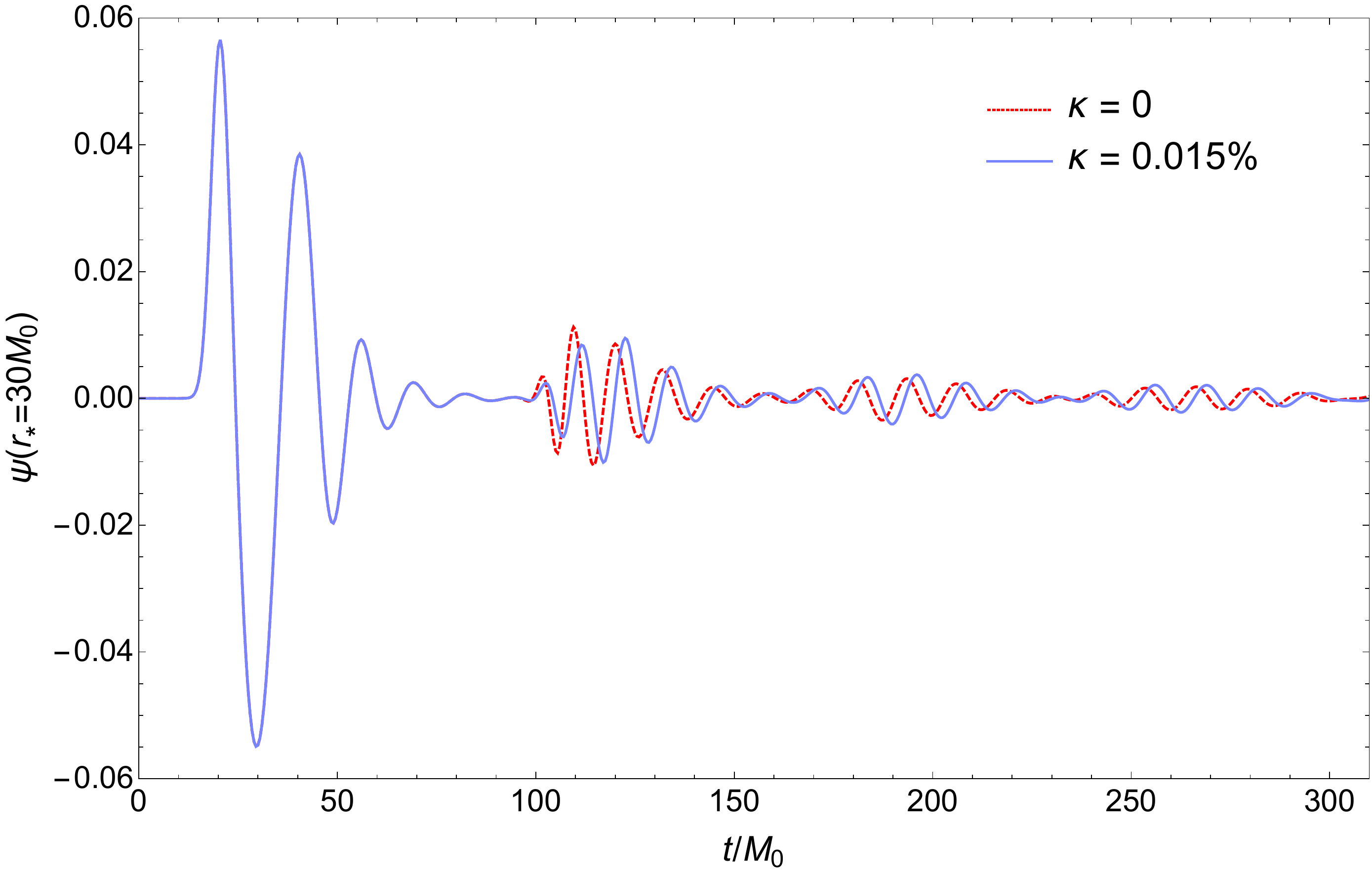}
\caption{Quadrupolar waveform extracted at $r_*=30 M_0$. The initial Gaussian pulse has energy $E=4.4 \cdot 10^{-3} M_0$ and it is scattered off an ECO with initial compactness parameter $\cmp=10^{-7}$, \ie\ $r_0=2.0000002 M_0$.
The red dashed line shows the case of a perfectly reflecting object, while the purple solid line shows the case in which the surface of the object absorbs incoming radiation with an absorption efficiency of $0.015\%$ and it expands to achieve the initial compactness in a finite amount of time.}
\label{retard}
\end{figure}

Alternatively, there might be a transient phase throughout the object has already absorbed energy but has not expanded enough to reach the original compactness.
The duration $\tau$ of the transient phase is crucial to determine the effect that it produces on the signal.
If $\tau$ is much smaller than the light crossing time between the surface and the potential, \ie\ $\tau \ll \Delta t_\text{echo}$, when the second echo arrives on the surface, the object will have already recovered its initial compactness and thus no visible shift in the time delay will be produced.
On the other hand, when $\tau > \Delta t_\text{echo}$, the time delay between the first and the second echo will be longer than in the perfectly reflecting case because the second echo will arrive on the surface when the compactness is still different from the initial one.
Then, while the object continues to expand, it also continues to absorb small amounts of energy from subsequent echoes, thus it will reach its initial compactness only when the energy of these echoes will become negligible.
However, we expect its compactness to become closer and closer to the initial one, and thus the time delay among these subsequent echoes to become smaller and smaller until it reaches the value corresponding to the initial compactness.
In \cref{retard} we plot the waveform for a Gaussian pulse off an ECO whose surface expands linearly, compared with a perfectly reflecting ECO.
In this example we have chosen the expansion rate $v$ such that the surface moves as $r_0(t)=r_0+ v \Delta t$ and the compactness is modelled as
\be
\cmp(t+\Delta t)=\frac{r_0+ v \Delta t}{2 M(t)+\kabs(\cmp(t)) \Delta E(t)}-1\,,
\ee
such that the initial compactness is reached after a transient phase $\tau\approx 65 M_0 \geq \Delta t_\text{echo}$ with respect to the beginning of the absorption. 

Before closing a final comment is in order.
In the previous scenario we have assumed a BH-like relation between the instantaneous radius and mass, \ie\ a direct proportionality $r_0(t)\propto M(t)$.
However, the relation between the radius and the mass of star-like compact objects can be more involved, implying that the compactness of these objects can be different for different values of the mass --- see \eg\ Fig.~1 of Ref.~\cite{Raposo:2018rjn}.
Nonetheless, we are here considering ultra-compact objects which are good BH mimickers and as such they are expected to be characterized by a BH-like behavior.
It is however worth mentioning that assuming $r_0(t)$ to be a more general function of $M(t)$ would imply that the expanding ECO would experience a change in its compactness parameter.
This, in turn, would lead to a variation of the time delay among echoes, and to a breaking of the aforementioned degeneracy with the perfectly reflecting case.

\section{Conclusions}

In this paper we have analyzed the response of an ECO against scalar perturbations to discuss possible effects of non-linear interactions on the echoes signal. 
In the first part, we have let the central compact object absorb part of the incoming radiation, resulting in an increase in its ADM mass and leading to changes in the spacetime in which the scalar field propagates.
The most important effect of this on the echo waveform is the loss of the quasi-periodicity of the signal.
In fact, the absorption of each echo changes the mass and thus the compactness of the object, and as a result, it increases the time delay among echoes which depends logarithmically on the compactness.
If the absorption continues without any other backreaction of the object, the time delay will continue to increase.
However, the energy of the $n$th echo is smaller than the previous ones, hence the change in the time delays becomes more and more negligible and it stabilizes to a constant value, unless the ECO collapses into a BH and no other echoes are produced.

In the second part we have considered some possible model-independent scenarios in which the object reacts to the absorption to prevent its collapse into a BH.
In the first scenario, the ECO can increase its compactness only up to a certain limiting value.
This can happen in at least two different ways: (i) the ECO absorption coefficient decreases for increasing compactness, going to zero when the asymptotic compactness is reached: (ii) the absorption coefficient is constant but once the asymptotic compactness is reached the object starts to expand in order to remain in the same equilibrium configuration.
The characteristic feature of this scenario is that the time delay between echoes gets longer and longer until it stabilizes to a constant value corresponding to the time delay of an object with the asymptotic compactness.
In the second scenario, the ECO expands instantaneously in order to prevent the collapse.
If the expansion rate maintains the same initial compactness, the resulting signal will have constant time delays among echoes like in the case of a perfectly reflecting ECO.
However, this type of expansion is an idealization and physically we expect that the original compactness is not recovered instantly but only after a transient phase.
When the transient phase is much shorter than the time delay between echoes, the effects on the waveform are negligible; while when it is comparable or greater than the time delay, the interval among echoes in the resulting signal is initially bigger than in the perfect reflecting case but then gets shorter, until it reaches the value corresponding to the time delay of the object at its initial compactness.
It might also be possible that the mass-radius relation for these objects is not linear as for BHs, and that stable configurations have different compactness for different values of the mass.
In this case, the object will not expand at constant compactness, breaking the degeneracy with the perfectly reflecting case.

At this point, one can wonder whether these changes in the time delay are actually detectable, or whether the signal can still be approximated as quasi-periodic.
First of all, we emphasize that in all cases analyzed here the change in the time delay is a feature of the first part of the signal.
The number of echoes interested by this effect depends on the initial compactness, the absorption coefficient and the damping factor between echoes, \ie\ the difference in amplitude among subsequent echoes.
Anyway, thanks to its bigger amplitude, it is the first part of the signal that is more likely detectable.

The time delay between the signal and the first echo is subjected to more uncertainties because it can be affected by non-linear physics during the merger.
Thus the observable for which the effect of absorption might be more important and non-negligible is usually the difference between the first two time delays: $\Delta t_2$ between the first and second echo and $\Delta t_3$ between the second and third echo.
As an example, consider an ECO with mass $M_0$, initial compactness $\cmp=10^{-5}$, an absorption coefficient $\kabs=0.06\%$ and assume that the energy carried by the first echo is $E_\text{echo} \approx 10^{-2} M_0$ with a damping factor between the first and second echo $\gamma=0.35$: the resulting relative difference in the time delays turns out to be $\Delta t_{12}=(\Delta t_2-\Delta t_1)/{\Delta t _2} \approx 5.7\%$.\footnote{Taking into account the interaction time $\Delta t_\text{int} \approx 4M_0$ during which the field travels inside the interior of the object, modifies the relative difference as $\Delta t_{12}\approx 5.3\%$.}
Note that in this example, since the total absorbed energy is $E_\text{tot}= \kabs \sum_{n=0}^{\infty} \gamma^n E_\text{echo} $, the hoop limit $r_0=2M_0$ is never reached. Obviously, the difference $\Delta t_{12}$ can be larger if we choose parameters that do not prevent the overcoming of the hoop limit after the absorption of some of the first echoes. 

Finally, in this work we have not considered the possible effects of a frequency-dependent absorption coefficient.
As subsequent echoes contains smaller frequencies to respect to the previous ones, if the ECO absorbs only higher frequencies than a given critical energy scale, like in the case of Boltzmann reflectivity~\cite{Oshita:2019sat}, subsequent echoes will be less absorbed and the absorption can even become negligible after some echoes.
However, excluding a possible expansion of the object, it seems that the above described mechanism would additionally require a non-trivial dependence of the critical energy scale on the compactness of the object in order to avoid the formation of a horizon. We leave this to future investigations.
 
In conclusion, we think that the possible phenomenology exposed by this investigation should be taken into account in future searches for echoes in ringdown signals.
In fact, the strategies adopted in these searches are usually based on the quasi-periodicity of the echoes signal~\cite{Abedi:2016hgu,Ren:2021xbe}, a feature that we showed can be partially lost in more realistic scenarios.
This seems to indicate that future generic searches for echoes, agnostic to any specific model of BH mimickers, should give way to more model-dependent analyses which would take into account the stability, or meta-stability, of such objects.
Last but not least, the relevance of such findings appears to deserve an extension to gravitational perturbations, and possibly even more relevantly, to rotating geometries.
We hope that this study will then stimulate further investigations in the theory and phenomenology of ECOs and help elaborate a more refined searching strategy for such objects.

\begin{acknowledgments}
We thank Enrico Barausse, Miguel Bezares and Paolo Pani for useful discussions.
The authors acknowledge funding from the Italian Ministry of Education and Scientific Research (MIUR) under the grant PRIN MIUR 2017-MB8AEZ\@.
\end{acknowledgments}

\appendix
\section{Numerical implementation\label{app:numerical}}

\begin{figure}[ht]
\centering
\includegraphics[height=0.35\textwidth]{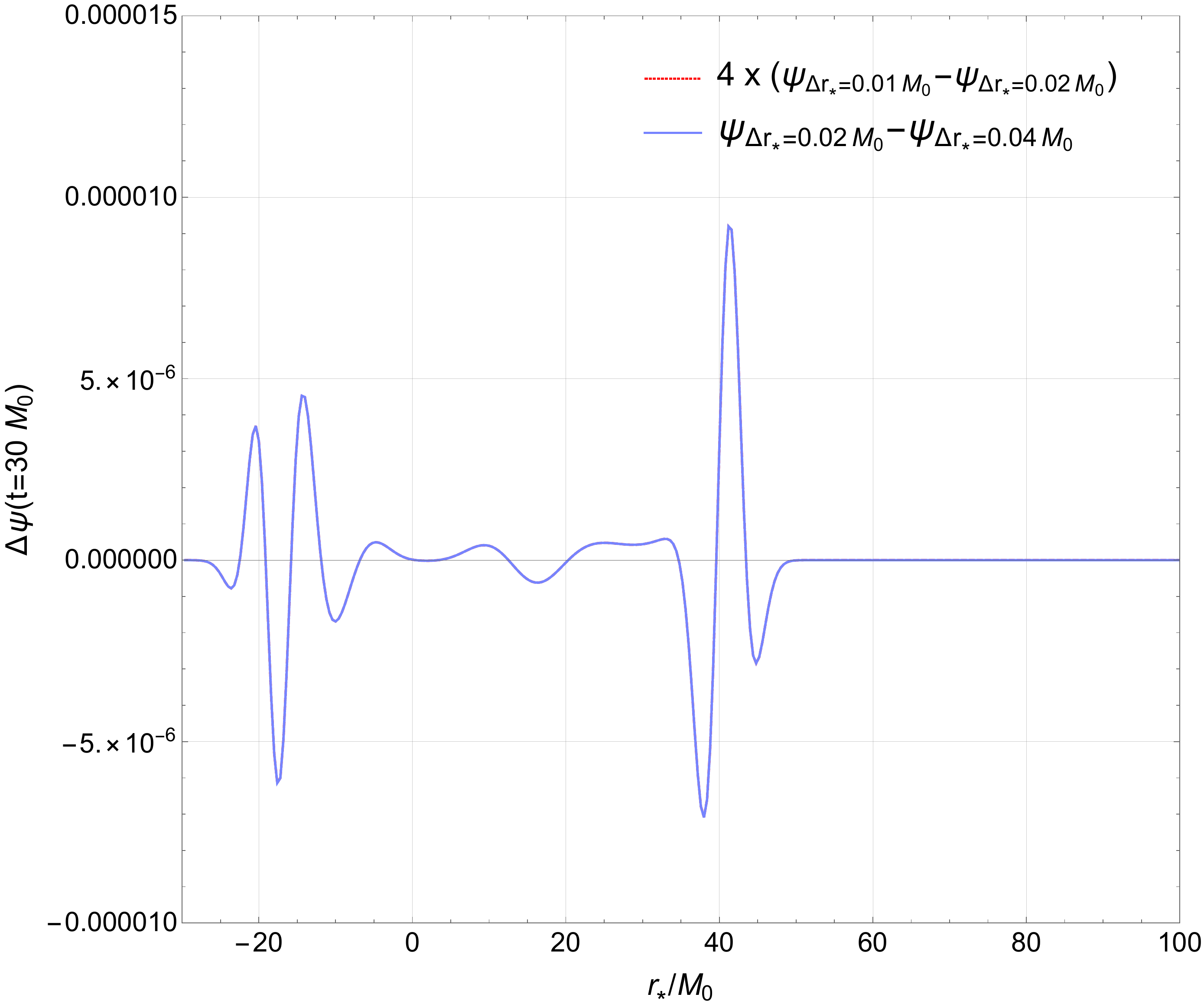}
\quad
\includegraphics[height=0.35\textwidth]{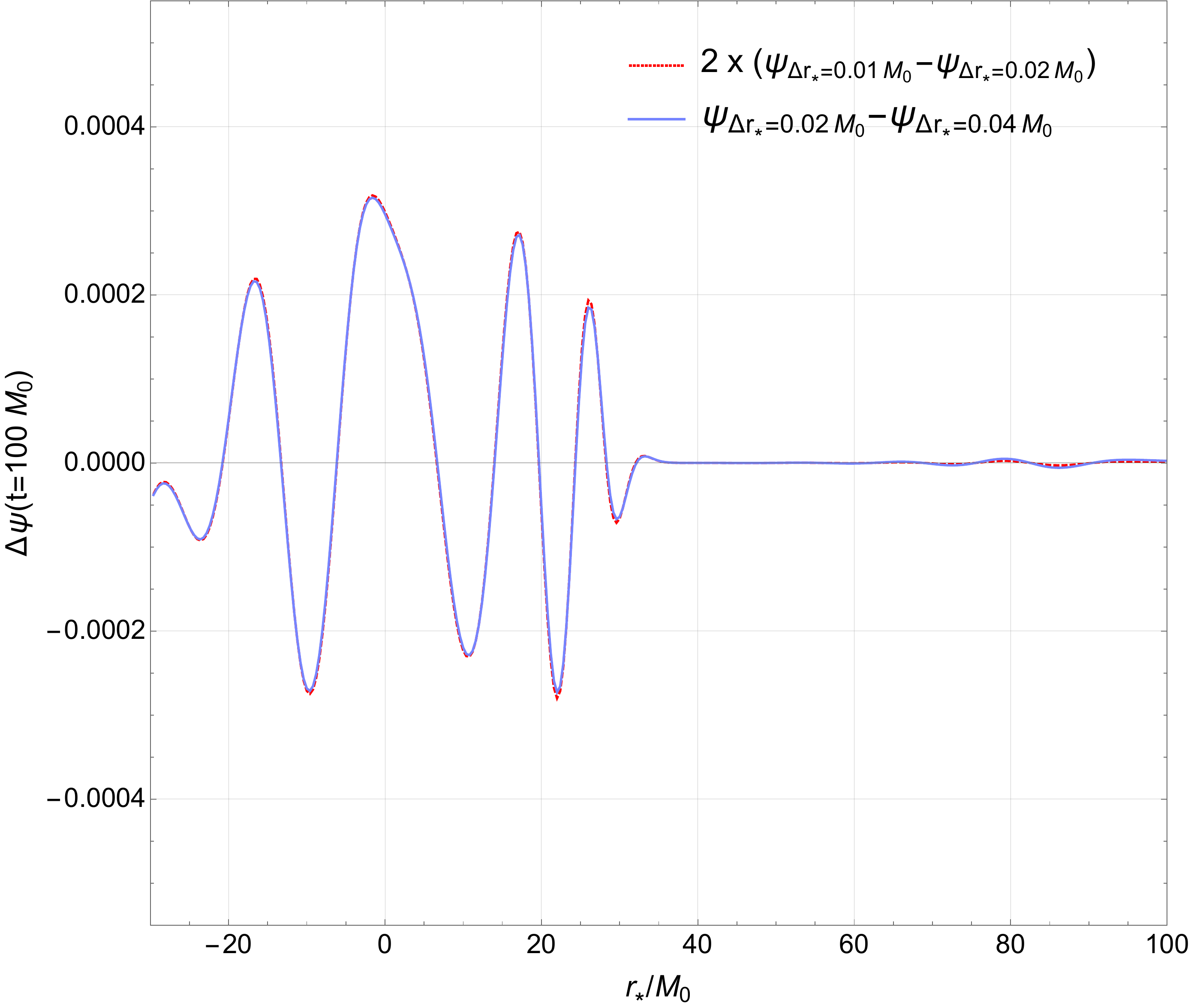}
\caption{Left: Convergence study of the evolution of the scalar field at time $t= 30 M_0$, when the mass parameter is still constant.
The purple line shows the difference between results obtained with low ($\Delta t= \Delta r_* = 0.04 M_0$) and medium ($\Delta t= \Delta r_* =0.02 M_0$) resolutions, while the red line shows the difference between results obtained with medium and high ($\Delta t= \Delta r_* =0.01 M_0$) resolutions multiplied by 4, the expected factor for second-order convergence.
Right: Convergence study of the evolution of the scalar field at time $t= 100 M_0$, when the mass is changed discretely at each time step, introducing a linear error.
The purple line shows the difference between results obtained with low ($\Delta t= \Delta r_* = 0.04 M_0$) and medium ($\Delta t= \Delta r_* =0.02 M_0$) resolutions, while the red line shows the difference between results obtained with medium and high ($\Delta t= \Delta r_* =0.01 M_0$) resolutions multiplied by 2, the expected factor for linear convergence.
In our simulations we always use higher resolution, \ie\ $\Delta t= \Delta r_* =0.001 M_0$.}
\label{fig:conv}
\end{figure}

In our simulations we have evolved the scalar field using a fourth-order Runge--Kutta integrator and computing spatial derivatives with finite-difference approximation of second-order accuracy~\cite{Dolan:2011dx}.
To validate our code we have checked the conservation of the energy and the matching with the time delays computed analytically.
We have also performed a convergence study to guarantee that the numerical resolution is high enough.
In the left panel of \cref{fig:conv} we observe that, until the mass of the central object is constant, we find the expected second order convergence.
However, the right panel of \cref{fig:conv} shows that as the field arrives at the surface and absorption is taken into account, the discrete \emph{linear} increase of the mass parameter at each time step introduces a dominant linear error. 

\section{Boundary conditions\label{app:boundary}}

The boundary conditions for the scenarios investigated in this work are non-trivial.
As usual, we impose purely outgoing (\ie\ perfectly absorbing) boundary conditions at the outer boundary.
This is because in real physical systems there is no outer boundary and the radiation escapes away at infinity.
To do so, we simply impose the condition
\be
\frac{\p \Psi}{\p t} +  \frac{\p \Psi}{\p x}=0
\ee
at the outer boundary of our computational domain.

However, differently from the BH case, at the ECO surface $r_0$ we need to impose \emph{partially} reflecting boundary conditions, to account for absorption.
%
One way to implement it is to insert a fictitious, dissipative region of length $l$ behind $r_0$ and then a perfectly reflecting boundary condition at $r_0-l$, so that looking at the reflected wave only from $r_0$ onwards it will have an effective smaller amplitude.
The dissipation, in turn, can be implemented through two possible methods. One is to switch on a dissipative term in this region through perfect matched layers~\cite{BERENGER1994185}.
The other method is to pause the simulation at an instant in which the whole part of the field that has to be reflected is present inside the fictitious region, then for each point of the region we replace $\Psi(r)$ with $(1-\kabs)^{1/2} \Psi(r)$ before the simulation stars over.
In any case, the fictitious dissipative region, will cause a delay in the reflection that can be either deleted in the final results or can be interpreted as the interaction time between the massive object and the scalar wave.
If the central object is compact enough, as in the cases considered here, this replacement can be done also without inserting any fictitious region, simply stopping the simulation when the part of the field that has to be reflected is in the region between the potential and the surface.

We also checked that we obtain the same results if we simply impose Dirichlet boundary conditions and multiply each $n$th echo for $(1-\kabs)^{n/2}$ at the end of the simulation.
We have used this last simple implementation for all the simulations presented here in which the absorption coefficient is constant.
This method has already been studied and applied even in the frequency domain \eg\ in Refs.~\cite{Cardoso:2019apo,Mark:2017dnq}.

Note also that, given the small absorption coefficients used in this work, the effect on the echoes amplitude is negligible.

\section{Absorbed energy\label{app:absorbed}}

During the multiple reflections of echoes, the central object absorbs part of the energy of the scalar field and its mass increases, at a given time step, by an amount $\kabs\,\Delta E(t)$, where $\kabs$ is the absorption coefficient defined in \cref{kabsorption} and $\Delta E(t)$ is the field energy present, at time $t$, in the last spatial bin of our computational domain corresponding to the location of the surface of the central object.

For this reason we need to know the energy distribution of the echoes that arrive at the surface. 
To obtain this distribution we run a simulation in which the central object is very compact, precisely we put the surface at $r_*=-70 M_0$. Then we look at the energy that pass through a fixed point, distant from the reflecting surface to avoid possible deformations and interference with the reflecting wave. Precisely we chose to evaluate the energy at $r_*=-45 M_0$. 
Since the field moves on light rays $\dd r_* = \dd t$, these energy distribution in time is equivalent to the spatial energy distribution.

\begin{figure}[ht]
\centering
\includegraphics[width=0.8\textwidth]{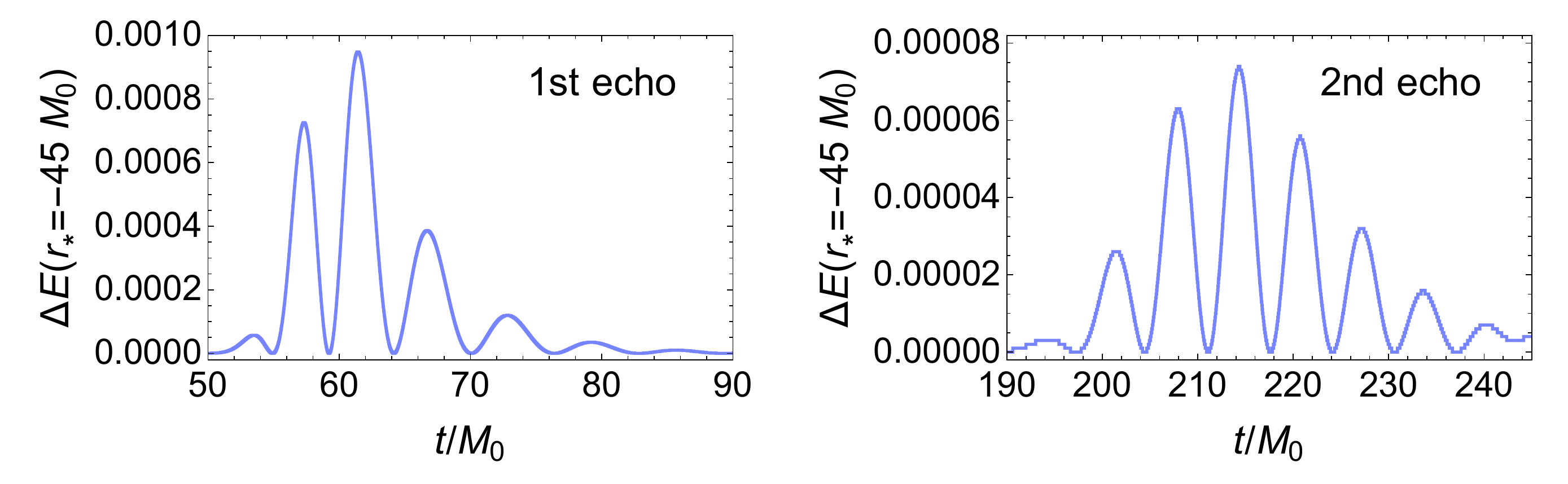}
\caption{Energy present in the spatial bin between $r_*=-45.001 M_0$ and $r_*=-45 M_0$ for the first two echoes of a quadrupolar Gaussian pulse in the spacetime of an ECO with compactness $\cmp=2.3 \cdot 10^{-16}$.}
\label{fig:dE}
\end{figure}

The results obtained in this way for the first two echoes are shown in \cref{fig:dE}. Taking into account the right time shift due to the different position of the surface in the true simulations, we obtain the energy that arrive at the object for each instant.

Note however that the results are poorly influenced by the precise distribution of the energy but basically depends only on the time of arrival of the echo, its spread and its total energy.

\bibliography{refs}

\end{document}